\documentclass[12pt,oneside]{article}

\usepackage{amssymb}
\usepackage{amsxtra}
\usepackage{amsmath}
\usepackage{amsfonts}
\usepackage{CJK}
\usepackage[titletoc]{appendix}
\usepackage{graphicx}

\numberwithin{equation}{section}

\newcommand{\eqa}{\begin{eqnarray}}
\newcommand{\eeqa}{\end{eqnarray}}
\newcommand{\beq}{\begin{equation}}
\newcommand{\eeq}{\end{equation}}
\newcommand{\nn}{\nonumber}
\newcommand{\p}{\partial}

\allowdisplaybreaks
\begin{document}

\date{}
\author{Beibei Hu$^{a,b,}$\thanks{Corresponding author. E-mail address: hu\_chzu@163.com(B.-b. Hu).} , Tiecheng Xia$^{b}$, Ling Zhang$^{a}$\\
\small \textit{a.School of Mathematicas and Finance, Chuzhou University, Anhui, 239000, China}\\
\small \textit{b.Department of Mathematics, Shanghai University, Shanghai 200444, China}}
\title{An integrable generalization of the super Kaup-Newell soliton hierarchy and its bi-Hamiltonian structure
}
\maketitle

\begin{abstract}

%% Text of abstract
An integrable generalization of the super Kaup-Newell(KN) isospectral problem is introduced and
its corresponding generalized super KN soliton hierarchy are established based on a Lie super-algebra
B(0,1) and super-trace identity in this paper. And the resulting super soliton hierarchy
can be put into a super bi-Hamiltonian form. In addition, a generalized super KN soliton hierarchy with
self-consistent sources is also presented.
\end{abstract}

\section{Introduction}

\quad\;\;As we all know, With the development of soliton theory, super integrable
systems associated with Lie super algebra have been
paid growing attention, many classical integrable equations
have been extended to be the super completely integrable equations [1-12]. Among those, Hu \cite{hu1997} and Ma \cite{ma2008} has made a
great contribution. Hu \cite {hu1997} proposed the super-trace identity, which is an effective
tool to constructing super Hamiltonian structures of super integrable equations.
In 2008, Ma given the proof of the super-trace identity and the super Hamiltonian structure
of many super integrable equations is established by the super-trace identity \cite{ma2008,ma2010}.

The soliton equation with self-consistent sources play an important role in discussing
the integrability for soliton hierarchy. They are relevant to some problems related to hydrodynamics,
solid state physics, plasma physics, and they are also usually used to describe
interactions between different solitary waves [16-18].
Very recently, self-consistent sources for super integrable equation hierarchy are constructed [19-30].

In Ref.\cite{Yan2002}, Yan considered a hierarchy of generalized KN equations,
where the spatial spectral problem is given by
\eqa {\small \phi_x=U\phi, U=\left(\begin{array}{cc}
\lambda+\mu qr&q\\
r&-\lambda-\mu qr
\end{array} \right), \phi=\left(\begin{array}{c}
\phi_1\\
\phi_2\end{array} \right), u=\left(\begin{array}{c}
q\\
r\end{array} \right),}\label{1.1}\eeqa
where $q$ and $r$ are both scalar potentials, $\lambda$ is the spectral parameter,
and $\mu$ is an arbitrary constant. The case of $\mu=0$ Eq.\eqref{1.1} reduces to the
well-know Kaup-Newell spectral problem \cite{Sirendoreji1998}, and $\mu=-\frac{1}{2}$
the Eq.\eqref{1.1} becomes the one considered by Qiao\cite{Qiao1994} by using the spectral
gradient method and nonlinearization approach. Another three versions of
generalized KN equations were also discussed in refs.[34-39]. The same idea
to generalize the AKNS hierarchy [40-42] and the Wadati-Konno-Ichikaw(WKI)
hierarchy \cite{Zhu2015}, whose bi-Hamiltonian structures were constructed.
Inspired by those generalizations, we would, in this paper, like to construct a generalized super KN hierarchy.

Organization of this paper. In Section 2, we shall construct a generalized super KN
hierarchy based on a Lie super-algebra. In Section 3, the super bi-Hamiltonian form
will be presented for the obtained generalized super KN hierarchy by making use of the super trace
identity and a generalized super KN hierarchy with self-consistent
sources generated from the variational derivative of spectra.
And some conclusions and discussions are given in the last Section.

\section{ A hierarchy of generalized super KN equations}

\quad\;\;In this section, we shall construct a generalized super KN hierarchy starting from a Lie super-algebra.
Consider the following spatial spectral problem
 \eqa {\small \phi_x=M\phi,
  M=\left(\begin{array}{ccc}
\lambda+\omega&q&\alpha\\
\lambda r&-\lambda-\omega&\lambda\beta\\
\lambda\beta&-\alpha&0
\end{array} \right),
\phi=\left(\begin{array}{c}
\phi_1\\
\phi_2\\
\phi_3\end{array} \right),u=\left(\begin{array}{c}
q\\
r\\
\alpha\\
\beta\end{array} \right),}\label{2.1}\eeqa
where $\omega=\mu(qr+2\alpha\beta)$ with $\mu$ is an arbitrary even constant,
$\lambda$ is the spectral parameter, $q$ and $r$ are even potentials,
and $\alpha$ and $\beta$ are odd potentials. Obviously, the spatial spectral problem \eqref{2.1}
with $\mu=0$ reduces to the standard super KN case \cite{Tao2011,wang2013}.

And associated with the Lie super-algebra
$G_1=\{\sum_{i=1}^5\lambda_ie_i|\lambda_i\in\mathcal{A},i=1,2,3,4,5\}$.
\eqa &&
e_1=\left(\begin{array}{ccc}
1&0&0\\
0&-1&0\\
0&0&0\end{array} \right),
e_2=\left(\begin{array}{ccc}
0&0&0\\
1&0&0\\
0&0&0\end{array} \right),
e_3=\left(\begin{array}{ccc}
0&1&0\\
0&0&0\\
0&0&0\end{array} \right),\nn\\
&&e_4=\left(\begin{array}{ccc}
0&0&1\\
0&0&0\\
0&-1&0\end{array} \right),
e_5=\left(\begin{array}{ccc}
0&0&0\\
0&0&1\\
1&0&0\end{array} \right).\nn\eeqa
which satisfy the following relationship
\eqa &&[e_1,e_2]=-2e_2,[e_1,e_3]=2e_3,[e_2,e_3]=-e_1,\nn\\
&&[e_5,e_1]=[e_2,e_4]=e_5,[e_3,e_4]=[e_2,e_5]=0,[e_3,e_5]=[e_1,e_4]=e_4,\nn\\
&&[e_4,e_4]_{+}=-2e_3,[e_5,e_5]_{+}=2e_2,[e_4,e_5]_{+}=[e_5,e_4]_{+}=e_1. \label{2.2}\eeqa
where $e_1,e_2,e_3$ are even and $e_4,e_5$ are odd, $[.,.]$ and $[.,.]_{+}$ denote the commutator and the anticommutator,
meanwhile $q,r$ are even elements and $\alpha,\beta$ are odd elements in \eqref{2.2}.

From the Tu format, We setting
\eqa {\small N=\left(\begin{array}{ccc}
A&B&\rho\\
\lambda C&-A&\lambda\delta\\
\lambda\delta&-\rho&0\end{array} \right)=\sum^n_{m=0}\left(\begin{array}{ccc}
a_m&b_m&\rho_m\\
\lambda c_m&-a_m&\lambda\delta_m\\
\lambda\delta_m&-\rho_m&0\end{array}
\right)\lambda^{-m} },\label{2.3}\eeqa
the corresponding $A,B,C$ are even elements and $\rho,\delta$ are odd elements, if we want to get the super integrable system,
we solve the stationary zero curvature equation at first
 \beq N_x=[M,N].\label{2.4}\eeq
Substituting $M$ in \eqref{2.1} and $N$ in \eqref{2.3} into Eq.\eqref{2.4} and comparing the coefficients of $\lambda^{-m} (m \geq 0)$,  we obtain
\eqa
\left\{\begin{array}{l}
a_{m,x}=qc_{m+1}-rb_{m+1}+\alpha\delta_{m+1}+\beta\rho_{m+1},\\
b_{m+1}=\frac{1}{2}b_{m,x}+qa_{m}+\alpha\rho_{m}-\omega b_{m},\\
c_{m+1}=-\frac{1}{2}c_{m,x}+ra_{m}+\beta\delta_{m+1}-\omega c_{m},\\
\rho_{m+1}=\rho_{m,x}+\alpha a_{m}+\beta b_{m+1}-q\delta_{m+1}-\omega \rho_{m},\\
\delta_{m+1}=-\delta_{m,x}+\beta a_{m}-\alpha c_{m}+r\rho_{m}-\omega \delta_{m}.\\
\end{array}\right.\label{2.5}\eeqa
which leads to a recursive relationship
\eqa
\left\{\begin{array}{l}
(c_{m+1}, b_{m+1},\delta_{m+1}, \rho_{m+1})^T=L( c_{m} ,b_{m},\delta_{m},\rho_{m})^T,\\
a_m=\p^{-1}(-\frac{1}{2}qc_{mx}-\frac{1}{2}rb_{mx}-\alpha\delta_{mx}+\beta\rho_{mx}-q\omega c_m+r\omega b_m-\alpha\omega\delta_m-\beta\omega\rho_m).\\
\end{array}\right.\label{2.6}\eeqa
Where the recursion operator $L$ has the following form
$$L=(L_{ij})_{4\times 4},\quad i,j=1,2,3,4,$$
with
\eqa &&
L_{11}=-\beta\alpha-(\omega+\frac{1}{2}\p)-r\p^{-1}q(\omega+\frac{1}{2}\p),\;\;
L_{12}=r\p^{-1}r(\omega-\frac{1}{2}\p),\nn\\&&
L_{13}=-r\p^{-1}\alpha(\p+\omega)-\beta(\p+\omega),
L_{14}=r\p^{-1}\beta(\p-\omega)+r\beta,
L_{21}=-q\p^{-1}q(\omega+\frac{1}{2}\p),\nn\\&&
L_{22}=q\p^{-1}r(\omega-\frac{1}{2}\p)+\frac{1}{2}\p-\omega,
L_{23}=-q\p^{-1}\alpha(\p+\omega),
L_{24}=q\p^{-1}\beta(\p-\omega)+\alpha,\nn\\&&
L_{31}=-\beta\p^{-1}q(\omega+\frac{1}{2}\p)-\alpha,
L_{32}=\beta\p^{-1}r(\omega-\frac{1}{2}\p),
L_{33}=-\beta\p^{-1}\alpha(\p+\omega)-(\p+\omega),\nn\\&&
L_{34}=\beta\p^{-1}\beta(\p-\omega)+r,
L_{41}=q\alpha-\alpha\p^{-1}q(\omega+\frac{1}{2}\p),
L_{42}=\alpha\p^{-1}r(\omega-\frac{1}{2}\p)-\beta(\omega-\frac{1}{2}\p),\nn\\&&
L_{43}=-\alpha\p^{-1}\alpha(\p+\omega)+q(\p+\omega),\;\;
L_{44}=\alpha\p^{-1}\beta(\p-\omega)+(\p-\omega)-qr.\label{2.7}\eeqa

For a given initial value $a_0=k_0\neq0,b_0=c_0=\rho_0=\delta_0=0$, the $a_j,b_j,c_j,\rho_j,\delta_j$$(j\geq1)$ can be
calculated by the recursion relation \eqref{2.6}. Here we list the several values
\eqa &&a_1=-\frac{1}{2}k_0(qr+2\alpha\beta),b_1=k_0q,c_1=k_0r,\rho_1=k_0\alpha,\delta_1=k_0\beta,\nn\\
&&a_2=k_0[\frac{3}{8}q^2r^2+\frac{3}{2}qr\alpha\beta+(qr+2\alpha\beta)\omega
+\frac{1}{4}(qr_x-q_xr)+(\alpha\beta_x-\alpha_x\beta)+\frac{3}{2}q\beta\beta_x],\nn\\
&&b_2=k_0[\frac{1}{2}q_x-\frac{1}{2}q(qr+2\alpha\beta)-q\omega],\;\;
c_2=-k_0[\frac{1}{2}r_{x}+\frac{1}{2}r(qr+2\alpha\beta)+r\omega+\beta\beta_x],\nn\\
&&\rho_2=k_0(\alpha_x-\frac{1}{2}\alpha qr+\frac{1}{2}\beta q_x+q\beta_x-\omega\alpha),\;\;
\delta_2=-k_0(\beta_x+\frac{1}{2}\beta qr+\omega\beta),\cdots\cdots. \nn\eeqa
Then, consider the auxiliary spectral problem associated with the spectral problem \eqref{2.1}
$$\phi_{t_n}=N^{(n)}\phi$$
where
\beq N^{(n)}=N_{+}^{(n)}+\Delta_n=\sum^n_{m=0}\left(\begin{array}{ccc}
a_m&b_m&\rho_m\\
\lambda c_m&-a_m&\lambda\delta_m\\
\lambda\delta_m&-\rho_m&0\end{array}
\right)\lambda^{n-m}
+\left(\begin{array}{ccc}
a&b&e\\
c&-a&f\\
f&-e&0\end{array}
\right), \label{2.8} \eeq
with $\Delta_n$ being the modification term, substituting Eq.\eqref{2.1} and Eq.\eqref{2.8} into the following zero curvature equation
\beq M_{t_n}-N_x^{(n)}+[M,N^{(n)}]=0, \label{2.9} \eeq
where $n\geq0$. Making use of Eq.\eqref{2.5}, we have
\eqa
\left\{\begin{array}{l}
\omega_{t_n}=a_x,b=c=e=f=0,\\
q_{t_n}=b_{n,x}+2qa_{n}+2\alpha\rho_{n}-2\omega b_{n}+2qa=2b_{n+1}+2qa,\\
r_{t_n}=c_{n,x}-2ra_{n}-2\beta\delta_{n+1}+2\omega c_{n}-2ra=-2c_{n+1}-2ra,\\
\alpha_{t_n}=\rho_{n,x}+\alpha a_{n}+\beta b_{n+1}-q\delta_{n+1}-\omega \rho_{n}+\alpha a=\rho_{n+1}+\alpha a,\\
\beta_{t_n}=\delta_{n,x}-\beta a_{n}+\alpha c_{n}-r\rho_{n}+\omega \delta_{n}-\beta a=-\delta_{n+1}-\beta a.\\
\end{array}\right.\label{2.10}\eeqa
which guarantees the following identity:
\beq (qr+2\alpha\beta)_{t_n}=-2(qc_{n+1}-rb_{n+1}+\alpha\delta_{n+1}+\beta\rho_{n+1})=-2a_{n,x}
\label{2.11} \eeq
Choosing $a=-2\mu a_n$, we can obtain the following hierarchy:
\beq
{\small u_{t_n}=\left(\begin{array}{c}
q\\
r\\
\alpha\\
\beta\end{array} \right)_{t_n}=\left(\begin{array}{c}
2b_{n+1}-4\mu qa_n\\
-2c_{n+1}+4\mu ra_n\\
\rho_{n+1}-2\mu\alpha a_n\\
-\delta_{n+1}+2\mu\beta a_n\end{array} \right)}. \label{2.12} \eeq
where $n\geq0$. The case of Eq.\eqref{2.12} with $\mu=0$ is exactly the super KN hierarchy \cite{Tao2011,wang2013}.
Therefore, Eq.\eqref{2.12} is called the generalized super KN hierarchy.

Taking $k_0=2,n=2$ in Eq.\eqref{2.12} and by using symbolic computation software (Maple), we obtain the first non-trivial flow is given by as follows:
 \beq  \left\{
\begin{array}{l}
q_{t_2}=q_{xx}-3qq_xr+\frac{3}{2}q^3r^2+4\alpha\alpha_x+8q\alpha\beta_x-4q\alpha_x\beta+6q^2\beta\beta_x+6q^2r\alpha\beta\\\quad\quad\;
+\mu(3q^3r^2-4q^2r_x-4qq_xr-8q_x\alpha\beta-8q\alpha\beta_x+8q\alpha_x\beta-12q^2\beta\beta_x\\ \quad\quad\;
+12q^2r\alpha\beta)-4\mu^2q^2r(qr+4\alpha\beta),\\
r_{t_2}=-r_{xx}-3qrr_x-\frac{3}{2}q^2r^3-6\beta\beta_{xx}-2qr^2\alpha\beta-10qr\beta\beta_x-4r\alpha\beta_x\\ \quad\quad\;
+\mu(-3q^2r^3-4q_xr^2-4qrr_x-8r_x\alpha\beta+8r\alpha\beta_x-8r\alpha_x\beta+4qr\beta\beta_x\\ \quad\quad\;
-4qr^2\alpha\beta)+4\mu^2qr^2(qr+4\alpha\beta),\\
\alpha_{t_2}=2\alpha_{xx}+2q_x\beta_x+\frac{3}{2}q_{xx}\beta-\frac{3}{2}q_xr\alpha
-\frac{3}{2}qr_x\alpha-2qr\alpha_x-\frac{3}{2}q^2r_x\beta-2q^2r\beta_x\\ \quad\quad\;
+\frac{3}{4}q^2r^2\alpha-3q\alpha\beta\beta_x-3qq_xr\beta+\mu(\frac{3}{2}q^2r^2\alpha-4q^2r\beta_x-2qr\alpha_x\\ \quad\quad\;
-3qr_x\alpha-q_xr\alpha-3q^2r_x\beta-6qq_xr\beta-14q\beta_x\alpha\beta)-2\mu^2q^2r^2\beta,\\
\beta_{t_2}=-2\beta_{xx}-r_x\alpha-2r\alpha_x-2qr\beta_x-\frac{3}{2}qr_x\beta-\frac{3}{2}q_xr\beta-\frac{3}{4}q^2r^2\beta+4\alpha\beta\beta_x\\ \quad\quad\;
+\mu(-3q_xr\beta-qr_x\beta-2qr\beta_x-\frac{3}{2}q^2r^2\beta)+2\mu^2q^2r^2\beta,
\end{array}
\right. \label{2.13} \eeq
whose Lax pair are $M$ in Eq.\eqref{2.1} and $N^{(2)}$ has the following form
\beq
N^{(2)}=\left(\begin{array}{ccc}
N_{11}^{(2)} & N_{12}^{(2)} & N_{13}^{(2)} \\
N_{21}^{(2)} & -N_{11}^{(2)} & N_{23}^{(2)} \\
N_{23}^{(2)} & -N_{13}^{(2)} &0
\end{array} \right).\label{2.14}\eeq
with
\eqa \left\{ \begin{array}{l}
N_{11}^{(2)}=2\lambda^2-\lambda(qr+2\alpha\beta)+(2-4\mu)[\frac{3}{8}q^2r^2+\frac{3}{2}qr\alpha\beta+(qr+2\alpha\beta)\omega
\\ \quad\quad\quad +\frac{1}{4}(qr_x-q_xr)+(\alpha\beta_x-\alpha_x\beta)+\frac{3}{2}q\beta\beta_x],\\
N_{12}^{(2)}=2q\lambda+q_x-q(qr+2\alpha\beta)-2q\omega,\\
N_{13}^{(2)}=2\lambda\alpha+2\alpha_x-qr\alpha+q_x\beta+2q\beta_x-2\omega\alpha,\\
N_{21}^{(2)}=2\lambda^2r-\lambda(r_x+r(qr+2\alpha\beta)+2r\omega+2\beta\beta_x),\\
N_{23}^{(2)}=2\lambda^2\beta-\lambda(2\beta_x+qr\beta+2\omega\beta).
\end{array}
\right. \nn\eeqa
when $\mu=\beta=\alpha=0$ and $t_2=t$, Eq.\eqref{2.13} just reduces to the well-know KN equation hierarchy\cite{Kaup1978}
\eqa \left\{ \begin{array}{l}
q_{t}=q_{xx}-3qq_xr+\frac{3}{2}q^3r^2 ,\\
r_{t}=-r_{xx}-3qrr_x-\frac{3}{2}q^2r^3 .
\end{array}
\right. \label{2.15} \eeqa

\section{Super bi-Hamiltonian structures}

\quad\;\;In what follows we shall find super bi-Hamiltonian structures of the generalized super KN hierarchy \eqref{2.12}.
To this end, we shall use the super trace identity, which proposed by Hu in \cite{hu1997}
and rigorously proved by Ma et al. in ref.\cite{ma2008}:
\beq \frac{\delta}{\delta u}\int Str(N\frac{\p M}{\p \lambda})dx=(\lambda^{-\gamma}\frac{\p}{\p\lambda}\lambda^{\gamma})Str(\frac{\p M}{\p u}N),\label{3.1}\eeq
where $Str$ denotes the super trace. It is not difficult to find that
\eqa &&
Str(N\frac{\p M}{\p \lambda})=2A+rB,
Str(\frac{\p M}{\p q}N)=2\mu rA+\lambda C,
Str(\frac{\p M}{\p r}N)=2\mu qA+\lambda B,\nn\\&&
Str(\frac{\p M}{\p \alpha}N)=4\mu\beta A+2\lambda\delta,
Str(\frac{\p M}{\p \beta}N)=4\mu\alpha A-2\lambda\rho,
\label{3.2}\eeqa
Substituting Eq.\eqref{3.2} into Eq.\eqref{3.1}, and comparing the coefficient of $\lambda^{-n-1}$ of both sides of Eq.\eqref{3.1} yields
\beq \left(\begin{array}{c}
\frac{\delta}{\delta q}\\
\frac{\delta}{\delta r}\\
\frac{\delta}{\delta \alpha}\\
\frac{\delta}{\delta \beta}\end{array}
\right)\int(2a_{n+1}+rb_{n+1})dx=(\gamma-n)\left(\begin{array}{c}
c_{n+1}+2\mu ra_{n}\\
b_{n+1}+2\mu qa_{n}\\
2\delta_{n+1}+4\mu\beta a_{n}\\
-2\rho_{n+1}+4\mu\alpha a_{n}\end{array} \right). \label{3.3}\eeq
To fix the vaule of $\gamma$, we let $n=0$ in Eq.\eqref{3.3} and find that (1) when $\mu=-\frac{1}{2}$,
$\gamma$ is arbitrary constant; (2) when $\mu\neq-\frac{1}{2}$, $\gamma=0$. Thus, we taking $\mu\neq-\frac{1}{2}$ yields
\beq
\left(\begin{array}{c}
c_{n+1}+2\mu ra_{n}\\
b_{n+1}+2\mu qa_{n}\\
2\delta_{n+1}+4\mu\beta a_{n}\\
-2\rho_{n+1}+4\mu\alpha a_{n}\end{array} \right)
=\frac{\delta \tilde{H}_n}{\delta u},\quad \tilde{H}_n=\int\frac{2a_{n+1}+rb_{n+1}}{n}dx. \label{3.4}\eeq
Moreover, it is easy to know that
\beq
\left(\begin{array}{c}
c_{n+1}\\
b_{n+1}\\
\delta_{n+1}\\
\rho_{n+1}\end{array} \right)
=R\left(\begin{array}{c}
c_{n+1}+2\mu ra_{n}\\
b_{n+1}+2\mu qa_{n}\\
2\delta_{n+1}+4\mu\beta a_{n}\\
-2\rho_{n+1}+4\mu\alpha a_{n}\end{array} \right)
\label{3.5}\eeq
where $R$ is given by
\beq
R=\left(\begin{array}{cccc}
1-2\mu r\p^{-1}q & 2\mu r\p^{-1}r & -\mu r\p^{-1}\alpha & \mu r\p^{-1}\beta\\
-2\mu q\p^{-1}q & 1+2\mu q\p^{-1}r & -\mu q\p^{-1}\alpha & \mu q\p^{-1}\beta\\
-2\mu \beta\p^{-1}q & 2\mu \beta\p^{-1}r & \frac{1}{2}-\mu \beta\p^{-1}\alpha & \mu \beta\p^{-1}\beta\\
-2\mu \alpha\p^{-1}q & 2\mu \alpha\p^{-1}r & -\mu \alpha\p^{-1}\alpha & -\frac{1}{2}+\mu \alpha\p^{-1}\beta
\end{array} \right)
\nn\eeq
Thus, the hierarchy of generalized super KN \eqref{2.12} possesses the following
super-Hamiltonian structure
\beq
{\small u_{t_n}=Q\left(\begin{array}{c}
c_{n+1}\\
b_{n+1}\\
\delta_{n+1}\\
\rho_{n+1}
\end{array} \right)=QR\left(\begin{array}{c}
c_{n+1}+2\mu ra_{n}\\
b_{n+1}+2\mu qa_{n}\\
2\delta_{n+1}+4\mu\beta a_{n}\\
-2\rho_{n+1}+4\mu\alpha a_{n}\end{array} \right)}=J\frac{\delta \tilde{H}_n}{\delta u}, n\geq1. \label{3.6} \eeq
where
\beq
Q=\left(\begin{array}{cccc}
-4\mu q\p^{-1}q & 2+4\mu q\p^{-1}r & -4\mu q\p^{-1}\alpha & -4\mu q\p^{-1}\beta\\
-2+4\mu r\p^{-1}q & -4\mu r\p^{-1}r & 4\mu r\p^{-1}\alpha & 4\mu r\p^{-1}\beta\\
-2\mu \alpha\p^{-1}q & 2\mu \alpha\p^{-1}r & -2\mu \alpha\p^{-1}\alpha & 1-2\mu \alpha\p^{-1}\beta\\
2\mu \beta\p^{-1}q & -2\mu \beta\p^{-1}r & -1+2\mu \beta\p^{-1}\alpha & 2\mu \beta\p^{-1}\beta
\end{array} \right)
\nn\eeq
and
\beq
J=QR=\left(\begin{array}{cccc}
-8\mu q\p^{-1}q & 2+8\mu q\p^{-1}r & -4\mu q\p^{-1}\alpha & 4\mu q\p^{-1}\beta\\
-2+8\mu r\p^{-1}q & -8\mu r\p^{-1}r & 4\mu r\p^{-1}\alpha & -4\mu r\p^{-1}\beta\\
-4\mu \alpha\p^{-1}q & 4\mu \alpha\p^{-1}r & -2\mu \alpha\p^{-1}\alpha & -\frac{1}{2}+2\mu \alpha\p^{-1}\beta\\
4\mu \beta\p^{-1}q & -4\mu \beta\p^{-1}r & -\frac{1}{2}+2\mu \beta\p^{-1}\alpha & -2\mu \beta\p^{-1}\beta
\end{array} \right)
\label{3.7}\eeq
here $J$ is a super Hamiltonian operator.

Specially, by making use of the recursive relationship \eqref{2.6},
the generalized super KN hierarchy \eqref{2.12} possesses the following
super-bi-Hamiltonian structure
\beq
{\small u_{t_n}=QL\left(\begin{array}{c}
c_{n}\\
b_{n}\\
\delta_{n}\\
\rho_{n}
\end{array} \right)=QLR\left(\begin{array}{c}
c_{n+1}+2\mu ra_{n}\\
b_{n+1}+2\mu qa_{n}\\
2\delta_{n+1}+4\mu\beta a_{n}\\
-2\rho_{n+1}+4\mu\alpha a_{n}\end{array} \right)}=P\frac{\delta \tilde{H}_{n-1}}{\delta u}, n\geq2. \label{3.8} \eeq
where the second compatible super-Hamiltonian operator $P=QLR=(P_{ij})_{4\times 4}, i,j=1,2,3,4,$ is given by
\eqa &&
P_{11}=4(\omega-\frac{1}{2}\p)\mu q\p^{-1}q+2q\Delta_1,
P_{12}=-4(\omega-\frac{1}{2}\p)\mu q\p^{-1}r-2(\omega-\frac{1}{2}\p)-2q\Delta_2,\nn\\&&
P_{13}=2(\omega-\frac{1}{2}\p)\mu q\p^{-1}\alpha+q\Delta_3,
P_{14}=-2(\omega-\frac{1}{2}\p)\mu q\p^{-1}\beta-\alpha-q\Delta_4,\nn\\&&
P_{21}=-4(\omega+\frac{1}{2}\p)\mu r\p^{-1}q-4\mu\beta\p\beta\p^{-1}q+2(\omega+\frac{1}{2}\p)-2\alpha\beta-2r\Delta_1\nn\\&&
P_{22}=4(\omega+\frac{1}{2}\p)\mu r\p^{-1}r+4\mu\beta\p\beta\p^{-1}r+2r\Delta_2,\nn\\&&
P_{23}=-2(\omega+\frac{1}{2}\p)\mu r\p^{-1}\alpha-2\mu\beta\p\beta\p^{-1}\alpha+\beta(\omega+\p)-r\Delta_3,\nn\\&&
P_{24}=2(\omega+\frac{1}{2}\p)\mu r\p^{-1}\beta+2\mu\beta\p\beta\p^{-1}\beta+r\beta+r\Delta_4,\nn\\&&
P_{31}=q\alpha+2\mu\omega\alpha\p^{-1}q-\mu\beta\p q\p^{-1}q-2\mu q\p\beta\p^{-1}q-2\mu\p\alpha\p^{-1}q+\alpha\Delta_1,\nn\\&&
P_{32}=-\beta(\omega-\frac{1}{2}\p)-2\mu\omega\alpha\p^{-1}r+\mu\beta\p q\p^{-1}r+2\mu q\p\beta\p^{-1}r+2\mu\p\alpha\p^{-1}r-\alpha\Delta_2,\nn\\&&
P_{33}=\frac{1}{2}q(\omega+\p)+\mu\omega\alpha\p^{-1}\alpha-\frac{1}{2}\mu\beta\p q\p^{-1}\alpha-\mu q\p\beta\p^{-1}\alpha-\mu\p\alpha\p^{-1}\alpha+\frac{1}{2}\alpha\Delta_3,\nn\\&&
P_{34}=\frac{1}{2}(\omega-\p)+\frac{1}{2}qr-\mu\omega\alpha\p^{-1}\beta+\frac{1}{2}\mu\beta\p q\p^{-1}\beta+\mu q\p\beta\p^{-1}\beta+\mu\p\alpha\p^{-1}\beta-\frac{1}{2}\alpha\Delta_4,\nn\\&&
P_{41}=-2\mu\p\beta\p^{-1}q-2\mu\omega\beta\p^{-1}q+\alpha-\beta\Delta_1,
P_{42}=2\mu\p\beta\p^{-1}r-2\mu\omega\beta\p^{-1}r-\beta\Delta_2,\nn\\&&
P_{43}=-\mu\p\beta\p^{-1}\alpha-\mu\omega\beta\p^{-1}\alpha-\frac{1}{2}\beta\Delta_3,
P_{44}=\mu\p\beta\p^{-1}\beta+\mu\omega\beta\p^{-1}\beta+\frac{1}{2}r-\frac{1}{2}\beta\Delta_4.\nn\eeqa
with
\eqa &&
\Delta_{1}=(2\mu-1)\p^{-1}q(\omega+\frac{1}{2}\p)-\mu(2\mu-1)\Delta\p^{-1}q,\nn\\&&
\Delta_{2}=(2\mu-1)\p^{-1}r(\omega-\frac{1}{2}\p)-\mu(2\mu-1)\Delta\p^{-1}r,\nn\\&&
\Delta_{3}=(2\mu-1)\p^{-1}\alpha(\omega+\p)-\mu(2\mu-1)\Delta\p^{-1}\alpha,\nn\\&&
\Delta_{4}=(2\mu-1)\p^{-1}\beta(\omega-\p)-\mu(2\mu-1)\Delta\p^{-1}\beta,\nn\eeqa
and
$$\quad \Delta=\p^{-1}q\p r+\p^{-1}r\p q+2\p^{-1}\alpha\p\beta-2\p^{-1}\beta\p\alpha.$$
Next, we are construct the generalized super KN hierarchy with self-consistent sources. At the super-isospectral problem
\eqa  \phi_x=M\phi,\quad \phi_t=N\phi. \label{3.9}\eeqa
Let $\lambda=\lambda_j$, the spectral vector corresponding $\phi$ remember to $\phi_j$, we obtain the the linear system as following
\eqa \left(\begin{array}{c}
\phi_{1j}\\
\phi_{2j}\\
\phi_{3j}\end{array} \right)_x=M_j\left(\begin{array}{c}
\phi_{1j}\\
\phi_{2j}\\
\phi_{3j}\end{array} \right),\quad\left(\begin{array}{c}
\phi_{1j}\\
\phi_{2j}\\
\phi_{3j}\end{array} \right)_t=N_j\left(\begin{array}{c}
\phi_{1j}\\
\phi_{2j}\\
\phi_{3j}\end{array} \right),\label{3.10}\eeqa
where $M_j=M|_{\lambda=\lambda_j}$, $N_j=N|_{\lambda=\lambda_j}$, $j=1,2...N$. By
\eqa  \frac{\delta \tilde{H}_n}{\delta
u}=\sum^N_{j=1}\frac{\delta \lambda_j}{\delta
u}=\sum^N_{j=1}\left(\begin{array}{c}
Str(\Psi_j\frac{\delta M}{\delta q})\\
Str(\Psi_j\frac{\delta M}{\delta r})\\
Str(\Psi_j\frac{\delta M}{\delta \alpha})\\
Str(\Psi_j\frac{\delta M}{\delta \beta})\end{array}\right)
=\left(\begin{array}{c}
<\Phi_2,\Phi_2>+2\mu r<\Phi_1,\Phi_2>\\
-<\Lambda\Phi_1,\Phi_1>+2\mu q<\Phi_1,\Phi_2>\\
-2<\Phi_2,\Phi_3>+4\mu\beta<\Phi_1,\Phi_2>\\
2<\Lambda\Phi_1,\Phi_3>+4\mu\alpha<\Phi_1,\Phi_2>
\end{array}\right),\label{3.11}\eeqa
where $\Phi_j=(\phi_{j1} ,\cdots,\phi_{jN})^T$, $j=1,2,3$.
So the generalized super KN hierarchy with self-consistent sources is proposed
\eqa
\small u_t=\left(\begin{array}{c}
q\\
r\\
\alpha\\
\beta\end{array} \right)_t=J\left(\begin{array}{c}
c_{n+1}+2\mu ra_{n}\\
b_{n+1}+2\mu qa_{n}\\
2\delta_{n+1}+4\mu\beta a_{n}\\
-2\rho_{n+1}+4\mu\alpha a_{n}\end{array} \right)+J\left(\begin{array}{c}
<\Phi_2,\Phi_2>+2\mu r<\Phi_1,\Phi_2>\\
-<\Lambda\Phi_1,\Phi_1>+2\mu q<\Phi_1,\Phi_2>\\
-2<\Phi_2,\Phi_3>+4\mu\beta<\Phi_1,\Phi_2>\\
2<\Lambda\Phi_1,\Phi_3>+4\mu\alpha<\Phi_1,\Phi_2>
\end{array}\right).\label{3.12} \eeqa
where $J$ is a super Hamiltonian operator given by in \eqref{3.7}.

\section{Conclusion and discussions}

\quad\;\;Starting from Lie super algebras, we may get super equation hierarchy.
With the help of variational identity, the Hamiltonian structure can also be presented.
Based on Lie super algebra, the self-consistent sources of a generalized super Kaup-Newell
hierarchy can be obtained. It enriched the content of self-consistent sources of super
soliton hierarchy. The methods in this study can be applied to other super soliton hierarchy to get
more super hierarchies with self-consistent will be discussed in our future work.\\
\\
\textbf{Acknowledgement}\\

This work is supported by the National Natural Science Foundation of China under Grant Nos. 11601055, 11271008 and 61072147.

\end{document}